\documentstyle[aps,prl,tighten,twocolumn]{revtex}
\input{psfig}

\begin{document}
\draft

\title{
Brightness of a phase-conjugating mirror behind a random medium}

\author{J. C. J. Paasschens,$^{\rm a,b}$
P.  W. Brouwer,$^{\rm a}$ and C. W. J. Beenakker$^{\rm a}$}
\address{$^a$Instituut-Lorentz, University of Leiden,
                 P.O. Box 9506, 2300 RA Leiden, The Netherlands\\
$^b$Philips Research Laboratories, 5656 AA Eindhoven, The
      Netherlands}
\date{\today}
\maketitle

\begin{abstract}
A random-matrix theory is presented for the reflection of light 
by a disordered medium backed by a phase-conjugating mirror.
Two regimes are distinguished, depending on the relative magnitude of
the inverse dwell time of a photon in the disordered
medium and the frequency shift acquired at the mirror.
The qualitatively different dependence of the reflectance
on the degree of disorder in the two regimes suggests a distinctive
experimental test for cancellation of phase shifts in a random medium.
\end{abstract}
\pacs{PACS numbers: 42.65.Hw, 42.25.Bs, 42.68.Ay, 78.20.Ci\\
 {\tt cond-mat/9608069}}
\narrowtext

A phase-conjugating mirror has the remarkable ability to cancel phase
shifts between incident and reflected
light\cite{Fisher,Zeldovich,Pep86}. This cancellation is used in optics
to correct for wave front distortions\cite{Wolf}: A plane wave which has been
distorted by an inhomogeneous medium is reflected at a phase-conjugating
mirror; After traversing the medium for a second time, the original
undistorted wave front is recovered. It is as if the reflected wave were
the time reverse of the incident wave.

Wave front correction is possible if the distorted wave front remains
approximately planar, because perfect time reversal upon reflection holds
only at a single angle of incidence. For this reason phase-conjugating
mirrors have been studied mainly in combination with weakly inhomogeneous
media. 
One might think that the diffusive illumination resulting from
a strongly inhomogeneous medium would render the effect of phase
conjugation insignificant. In this paper we show that, on the contrary,
phase conjugation drastically modifies the reflectance even if 
propagation through the medium is completely diffusive.

\begin{figure}
\centerline{
\psfig{file=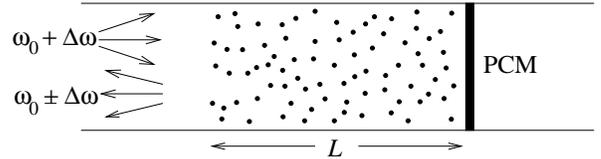,width=0.9\hsize}}
\medskip
\caption{Schematic drawing of the disordered medium backed by a
phase-conjugating mirror (PCM). Light incident at frequency
$\omega_0+\Delta\omega$ is reflected at the two frequencies
$\omega_0\pm\Delta\omega$.}
\label{figsys}
\end{figure}

The phase-conjugating mirror we consider consists of a material with a
large non-linear susceptibility, which is pumped by two
counter-propagating beams at frequency~$\omega_0$. A disordered medium
(length $L$, mean free path~$l$) is placed in front of the mirror, and
illuminated at the other end by a diffusive source at frequency
$\omega_0+\Delta\omega$ (see Fig.~\ref{figsys}). The reflected light
contains both frequencies $\omega_0+\Delta\omega$ and
$\omega_0-\Delta\omega$. The reflectances $R_+$, $R_-$
 are defined as the reflected power at frequency
$\omega_\pm\equiv\omega_0\pm\Delta\omega$ divided by the incident 
power. 
(Note the difference between the system we study here and the system studied in
e.g.\ Ref.~\cite{Krav90}, where phase conjugation occurs inside the
random medium itself.)
We distinguish a degenerate and a non-degenerate regime, 
depending on the relative magnitude of the
frequency shift $\Delta\omega$ and the inverse of the mean dwell time
$\tau_{\rm dwell}\simeq L^2/cl$ of a photon (velocity $c$) in the
disordered medium. In the degenerate regime, $\Delta\omega\ll1/\tau_{\rm
dwell}$, phase conjugation leads to constructive interference of multiply
scattered light in the disordered medium. In the non-degenerate regime,
$\Delta\omega\gg1/\tau_{\rm
dwell}$, interference effects are insignificant. A distinguishing
feature of the two regimes is that the reflectance $R_-$ decreases
monotonically as a function of $L/l$ in the degenerate regime, while in
the non-degenerate regime it first decreases and then increases.
The disappearance of the reflectance minimum 
on reducing $\Delta\omega$, provides a distinctive experimental test
for phase-shift cancellation in a random
medium.

Our theoretical approach applies techniques of random-matrix
theory\cite{Sto91} which were originally developed for the study of 
phase conjugation of electrons by a superconductor\cite{sup}.
To reduce the optical problem to the scattering of a scalar wave, we
choose a two-dimensional geometry. The scatterers consist of dielectric
rods in the $z$-direction, randomly placed in the $x$-$y$-plane. The
electric field points in the $z$-direction and varies in the
$x$-$y$-plane only. Two-dimensional scatterers are somewhat artificial,
but can be realized experimentally\cite{Freund88}. We believe that our
results apply qualitatively to a three-dimensional geometry as well,
because the randomization of the polarization by the disorder renders
the vector character of the light insignificant.

The $z$-component of the electric field at the frequencies 
$\omega_\pm$ is given by
\begin{equation}
  E_\pm(x,y,t) = {\rm Re}\, {\cal E}_\pm(x,y) \exp(-i\omega_\pm t).
\end{equation}
The phase-conjugating mirror (at $x=0$) couples the two frequencies via
the wave equation\cite{Fisher,Lenstra,Hout91}
\begin{equation}
  \pmatrix{{\cal H}_0&\gamma^*\cr-\gamma&-{\cal H}_0}\pmatrix{{\cal E}_+\cr
   {\cal E}_-^*} =
  \frac{2\varepsilon\Delta\omega}{\omega_0}\,\pmatrix{{\cal E}_+\cr 
   {\cal E}_-^*}.
   \label{BdGlike}
\end{equation}
The complex dimensionless coupling constant $\gamma$ is zero for $x<0$
and for $x>L_c$, with $L_c$ the length of the non-linear medium
forming the phase-conjugating mirror. For $0<x<L_c$ we put
$\gamma=\gamma_0 e^{i\psi}$.
The Helmholtz operator ${\cal H}_0$ at frequency $\omega_0$ is given 
by\cite{voetnoot}
\begin{equation}
   {\cal H}_0 =
   -k_0^{-2} \nabla^2 -\varepsilon,
  \label{Hdef}
\end{equation}
where $k_0 = \omega_0/c$ and  $\varepsilon(x,y)$ is the relative 
dielectric constant of the
medium. We take $\varepsilon=1$ everywhere 
except in the disordered region $-L<x<0$,
where $\varepsilon = 1 + \delta\varepsilon(x,y)$. The spatial fluctuations
$\delta\varepsilon$ lead to scattering with mean free path $l\gg1/k_0$.
The validity of Eq.~(\ref{BdGlike}) requires $\Delta\omega/\omega_0\ll1$
and $\gamma_0 \ll 1$. The ratio of these two small
parameters
\begin{equation}
  \delta\equiv {2\Delta\omega}/{\gamma_0\omega_0}
  \label{deltadef}
\end{equation}
is arbitrary.

To define finite-dimensional scattering matrices we embed the disordered
medium in a waveguide (width~$W$), containing $N_\pm = {\rm Int}
(\omega_\pm W/c\pi) \gg 1$ propagating modes at frequency $\omega_\pm$.
A basis of scattering states consists of the complex fields
\begin{mathletters}
  \label{basisdef}
\begin{eqnarray}
  E_{\pm,n}^> &=& k_{\pm,n}^{-1/2}\, \sin\left(\frac{n\pi y}{W}\right)\;
       \exp(ik_{\pm,n} x-i\omega_\pm t),\\
  E_{\pm,n}^< &=& k_{\pm,n}^{-1/2}\, \sin\left(\frac{n\pi y}{W}\right)\;
       \exp(-ik_{\pm,n} x-i\omega_\pm t).
\end{eqnarray}
\end{mathletters}%
Here $n=1,2,\ldots, N_\pm$ is the mode index and the superscript $>$
($<$) indicates a wave moving to the right (left), with frequency
$\omega_\pm $ and wavenumber $k_{\pm,n}$.
The normalization in Eq.~(\ref{basisdef}) is such that each
wave carries the same flux.

With respect to this basis, incoming and outgoing waves
are decomposed as
\begin{mathletters}
\begin{eqnarray}
  E^{\rm in} &=& \sum_{n=1}^{N_+} u_{+,n} E_{+,n}^> +
                 \sum_{n=1}^{N_-} u_{-,n} E_{-,n}^>, \\
  E^{\rm out}&=& \sum_{n=1}^{N_+} v_{+,n} E_{+,n}^< +
                 \sum_{n=1}^{N_-} v_{-,n} E_{-,n}^< .
\end{eqnarray}
\end{mathletters}%
The complex coefficients are combined into two vectors
\begin{mathletters}
\label{uvdef}
\begin{eqnarray}
  {\bf u} &=& (u  _{+,1}, u  _{+,2},\ldots, u  _{+,N_+},
         u^*_{-,1}, u^*_{-,2},\ldots, u^*_{-,N_-})^{\rm T},\\
  {\bf v} &=& (v  _{+,1}, v  _{+,2},\ldots, v  _{+,N_+},
         v^*_{-,1}, v^*_{-,2},\ldots, v^*_{-,N_-})^{\rm T}.
\end{eqnarray}
\end{mathletters}%
The reflection matrix ${\bf r}$ relates ${\bf u}$ to ${\bf v}$,
\begin{equation}
  {\bf v} = {\bf r}{\bf u},\qquad {\bf r}=
  \pmatrix{{\bf r}_{++} & {\bf r}_{+-}\cr {\bf r}_{-+}&{\bf r}_{--}}.
  \label{Sdef}
\end{equation}
The submatrices
${\bf r}_{\pm\pm}$ have dimensions $N_\pm\times N_\pm$. 
The reflectances $R_\pm$ are defined by
\begin{equation}
  R_- = N_+^{-1} \mathop{\rm Tr} {\bf r}_{-+}^{\vphantom \dagger}
            {\bf r}_{-+}^\dagger,\quad
  R_+ = N_+^{-1} \mathop{\rm Tr} {\bf r}_{++}^{\vphantom \dagger}
            {\bf r}_{++}^\dagger.
\label{R-R+def}
\end{equation}
For $\Delta\omega\ll\omega_0$ we may neglect the difference between $N_+$
and $N_-$ and replace both by $N={\rm Int} (k_0 W/\pi)$.

We construct ${\bf r}$ from the reflection matrix
${\bf r}_{\rm PCM}$ of the phase-conjugating mirror and the scattering 
matrix ${\bf S}_{\rm }$ of the disordered medium.
In the absence of disorder, an incoming plane wave in the direction
$(\cos\phi,\sin\phi)$ is retro-reflected at the phase-conjugating mirror
in the direction
$(-\cos\phi,-\sin\phi)$, with a different frequency and amplitude. The
$2N\times2N$
reflection matrix ${\bf r}_{\rm PCM}$ is\cite{Fisher,Lenstra,Hout91}
\begin{mathletters}
\label{adeffull}
\begin{eqnarray}
  {\bf r}_{\rm PCM} &=& \pmatrix{0&-i{\bf a} e^{-i\psi}\cr
                           i{\bf a} e^{i\psi}&0},\\
  a_{nm} &=& \delta_{nm} a(\phi_n),\qquad
  \phi_n = \arcsin(n\pi/k_0W),\\
  a(\phi) &=& \biggl[\sqrt{1+\delta^2}\mathop{\rm cotan}
    \biggl(\frac{\alpha\sqrt{1+\delta^2}}{\cos\phi}\biggr) 
     +i\delta\biggr]^{-1},
\end{eqnarray}
\end{mathletters}%
with $\alpha \equiv \case12 \gamma_0 k_0 L_c$.
The disordered medium in front of the phase-conjugating mirror does not
couple $\omega_+$ and~$\omega_-$. Its scattering properties at frequency
$\omega$ are described by two $N\times N$ transmission matrices
${\bf t}_{21}(\omega)$ and ${\bf t}_{12}(\omega)$ 
(transmission from left to right and from right to left),
plus two $N\times N$ reflection matrices
${\bf r}_{11}(\omega)$ and ${\bf r}_{22}(\omega)$ (reflection from left to left and
from right to right). Taken together, these four matrices constitute a
$2N\times 2N$ scattering matrix 
\begin{equation}
  {\bf S}_{\rm}(\omega) = 
  \pmatrix{ {\bf r}_{11}(\omega) & {\bf t}_{12}(\omega) \cr
            {\bf t}_{21}(\omega) & {\bf r}_{22}(\omega) },
\end{equation}
which is unitary (because of flux conservation) and symmetric (because of
time-reversal invariance).
(In contrast, ${\bf r}_{\rm PCM}$ is not flux conserving.)
Without loss of generality the reflection and transmission matrices of
the disordered region can be decomposed as \cite{Sto91}
\begin{mathletters}
  \label{tdef}
\begin{eqnarray}
  {\bf r}_{11}(\omega_\pm) &=& i{\bf U}\!_\pm^{\vphantom T} 
       \sqrt{{\mbox{\boldmath$\rho$}}_\pm}\, {\bf U}_\pm^{\rm T},\;
  {\bf t}_{21}(\omega_\pm)  =   {\bf V}\!_\pm^{\vphantom T} 
       \sqrt{{\mbox{\boldmath$\tau$}}_\pm}\, {\bf U}_\pm^{\rm T},\\
  {\bf r}_{22}(\omega_\pm) &=& i{\bf V}\!_\pm^{\vphantom T} 
       \sqrt{{\mbox{\boldmath$\rho$}}_\pm}\, {\bf V}_\pm^{\rm T},\;
  {\bf t}_{12}(\omega_\pm)  =   {\bf U}\!_\pm^{\vphantom T} 
       \sqrt{{\mbox{\boldmath$\tau$}}_\pm}\, {\bf V}_\pm^{\rm T}.
\end{eqnarray}
\end{mathletters}%
Here ${\bf U}_\pm$ and ${\bf V}_\pm$ are  $N\times N$ unitary
matrices, and $\mbox{\boldmath$\tau$}_\pm\equiv1-
\mbox{\boldmath$\rho$}_\pm$ is a diagonal 
matrix with the 
transmission eigenvalues $T_{\pm,n}\in[0,1]$ on the diagonal. 

Combining Eqs.~(\ref{R-R+def})--(\ref{tdef}) we find expressions for
$R_\pm$ in terms of $\mbox{\boldmath$\tau$}_\pm$ and 
${\bf\Omega} = {\bf V}_-^\dagger {\bf a}{\bf V}_+$. The
expression for $R_-$ is
\label{reflect}
\begin{eqnarray}
  R_- &=& N^{-1} \mathop{\rm Tr}
    \mbox{\boldmath$\tau$}_- {\bf \Omega}
   \left(1-\sqrt{\mbox{\boldmath$\rho$}_+}\,{\bf\Omega}^{\rm T}
           \sqrt{\mbox{\boldmath$\rho$}_-}\,{\bf\Omega}\right)^{-1}
   \mbox{\boldmath$\tau$}_+\nonumber\\
   &&\qquad\mbox{}\cdot
   \left(1-{\bf \Omega}^\dagger\sqrt{\mbox{\boldmath$\rho$}_-}\,
           {\bf \Omega}^*      \sqrt{\mbox{\boldmath$\rho$}_+}\right)^{-1}
   {\bf \Omega}^\dagger.
  \label{R-inmat}
\end{eqnarray}
The expression for $R_+$ is similar (but more lengthy).
To compute the averages $\langle R_\pm\rangle$ we have to average over
$\mbox{\boldmath$\tau$}_\pm$ and ${\bf V}_\pm$.
We make the isotropy approximation\cite{Sto91}
that the matrices ${\bf V}_\pm$ are uniformly distributed over the unitary
group ${\cal U}(N)$. 
For $\tau_{\rm dwell}\Delta\omega\ll 1$ we may
identify ${\bf V}_+={\bf V}_-$, while
for $\tau_{\rm dwell}\Delta\omega\gg 1$ the matrices
${\bf V}_+$ and ${\bf V}_-$ are 
independent. In each case the average 
over ${\cal U}(N)$ with $N\gg1$ can be done using the large
$N$-expansion of Ref.~\cite{Bro96}. The remaining average
over $T_{\pm,n}$ can be done using the known 
density $\rho(T)$ of the transmission eigenvalues in a disordered
medium\cite{Sto91}.

In the non-degenerate regime ($\tau_{\rm dwell}\Delta\omega\gg 1$)
the result is 
\begin{mathletters}
\label{radiattrans}
\begin{eqnarray}
  \langle R_-\rangle &=& \frac{T_0^2 A}{1-(1-T_0)^2 A^2}, \\
  \langle R_+\rangle &=& 1-T_0 +\frac{T_0^2(1-T_0) A^2}{1-(1-T_0)^2 A^2} ,
\end{eqnarray}
\end{mathletters}%
where $T_0 = (1+2L/\pi l)^{-1}$ is the transmittance at frequency
$\omega_0$ of the disordered medium in the large-$N$ limit\cite{Jong94}. 
The quantity $A=N^{-1}\mathop{\rm Tr} {\bf a}{\bf a}^\dagger$ is the 
modal average of the reflectance of the phase conjugating mirror
($A\to\int_0^{\pi/2}d\phi\,|a(\phi)|^2 \cos\phi$ for $N\to\infty$).
Eq.~(\ref{radiattrans}) can also be obtained within the framework of
radiative transfer theory, in which interference effects in the
disordered medium are disregarded\cite{Paa96}.

\begin{figure}
\centerline{
\psfig{file=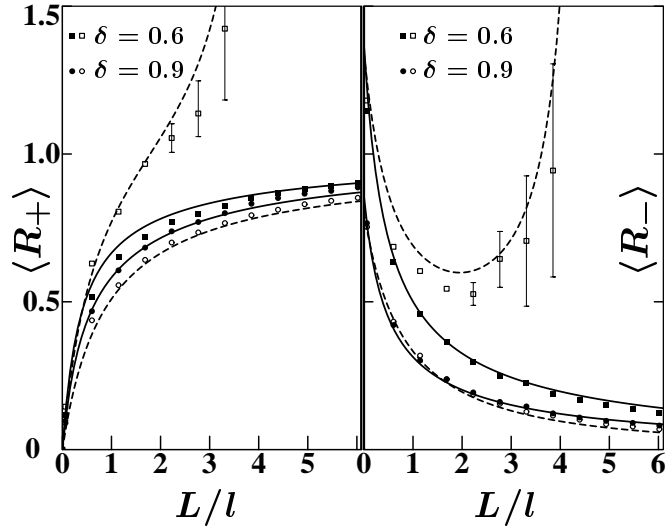,width=1.0\hsize}}
\medskip
\caption{Average reflectances $\langle R_\pm\rangle$ as a function of
$L/l$ for
$\alpha=\pi/4$ and $\delta=0.6$, 0.9.
The dashed curves are the non-degenerate case, given by
Eq.~(\protect\ref{radiattrans}).
The solid curves are the degenerate case, given by
Eq.~(\protect\ref{R-simpl}) for $L/l\protect\gtrsim 3$.
Data points are results from numerical simulations (open symbols for
the
non-degenerate case, filled symbols for the degenerate case).
Error bars are the statistical uncertainty of the average over 150
disorder configurations. (When the error bar is not shown it is smaller
than the size of the marker.)}
\label{figsimul}
\end{figure}

In Fig.~\ref{figsimul} we have plotted the result~(\ref{radiattrans})
for  $\langle R_\pm\rangle$ in the non-degenerate regime, as a function
of $L/l$ for $\alpha=\pi/4$ and two choices of $\delta=0.6$, $0.9$
(dashed curves). For $A>1$ (corresponding to $\delta=0.6$) the
reflectance $\langle R_-\rangle$ has a minimum at $L/l =
\case12\pi(A^2-1)^{-1}$, and both $\langle R_+\rangle$ and $\langle
R_-\rangle$ diverge at $L/l=\case12\pi(A-1)^{-1}$.
This divergence is preempted by depletion of the pump beams in
the phase-conjugating mirror, and signals the
breakdown of a stationary
solution to the scattering problem.
For $A<1$ (corresponding to $\delta=0.9$) $\langle R_-\rangle$ tends
to 0 as
$L^{-2}$ for $L\to\infty$, while $\langle R_+\rangle$ approaches 1 as
$L^{-1}$.

The situation is entirely different in the degenerate regime ($\tau_{\rm
dwell}\Delta\omega\ll 1$). 
The complete result is a complicated function of $L/l$ (plotted in
Fig.~\ref{figsimul}, solid curves).
For $L/l\to\infty$ the result takes the simpler form 
\begin{mathletters}
  \label{R-simpl}
\begin{eqnarray}
  \langle R_-\rangle &=& 2T_0\,
   {\rm Re} \frac{a_0^*(a_0^2-1)}{a_0^2-a_0^{*2}}\mathop{\rm arctanh} a_0, \\
  \langle R_+\rangle
   &=& 1 -2T_0\, {\rm Re} \frac{a_0^*(a_0^2-1)} {a_0^2-a_0^{*2}} 
      \mathop{\rm arctanh} a_0^*,
\end{eqnarray}
where the complex number $a_0$ is determined by
\begin{equation}
  \int_0^{\pi/2} d\phi\,\frac{\cos\phi\; a(\phi)}{1-a_0\,a(\phi)} =
       \frac{a_0}{1-a_0^2}.
  \label{a0implicitint}
\end{equation}
\end{mathletters}%
When $\delta\to0$,  $a_0\to1.284 - 0.0133\,i$ for $\alpha=\pi/4$. 
Both $\langle R_-\rangle$ and $\langle R_+\rangle$ have a monotonic 
$L$-dependence, tending to $0$ and $1$, respectively, as $1/L$ for
$L\to\infty$.

To test the analytical predictions of random-matrix theory  we
have carried out numerical simulations. The Helmholtz equation
\begin{equation}
  \left(\nabla^2 + \varepsilon\omega_\pm^2/c^2\right) {\cal E} =0
\end{equation}
is discretized on a square lattice (lattice constant $d$, 
length $L$, width $W$). 
The relative dielectric constant $\varepsilon$ fluctuates from site to
site between $1\pm\delta\varepsilon$. 
Using the method of recursive Green functions\cite{Bar91} we compute the
scattering matrix ${\bf S}_{\rm}(\omega)$ of the disordered medium at
frequencies $\omega_+$ and $\omega_-$.
The reflection matrix ${\bf r}_{\rm PCM}$ of the phase-conjugating mirror is
calculated by discretizing Eq.~(\ref{BdGlike}). From  ${\bf S}
(\omega_\pm)$ and ${\bf r}_{\rm PCM}$ we obtain the reflection matrix
${\bf r}$ of the entire system, and hence the reflectances $R_\pm$ using
Eq.~(\ref{R-R+def})

We took $W=51\,d$, $\delta\varepsilon=0.5$, $\alpha=\pi/4$, and varied
$\delta$ and $L$. For the degenerate case we took $\omega_+=\omega_- =
1.252\,c/d$, and for the non-degenerate case $\omega_+=1.252\,c/d$,
$\omega_-=1.166\,c/d$. These parameters correspond to $N_+=22$,
$l_+=15.5\,d$ at frequency $\omega_+$. (The mean free path is determined
from the transmittance of the disordered region.)
In the non-degenerate case we have $N_-=20$,
$l_-=20.1\, d$. For comparison with the analytic theory, where the
difference between $N_+$ and $N_-$ and between $l_+$ and $l_-$ is
neglected, we use the values $N_+$ and $l_+$.
Results for the average reflectances (averaged over 150 realizations of
the disorder) are shown in Fig.~\ref{figsimul}, and are in good 
agreement with the analytical predictions.

A striking feature of the degenerate regime is the absence of the minimum 
in $\langle R_-\rangle$ as a function of $L/l$ for $A>1$.
A qualitative explanation for the disappearance of the reflectance
minimum goes as follows.
To first order in $L/l$, disorder reduces the intensity of light reflected with
frequency shift $2\Delta\omega$, because some light is scattered back
before it can reach the phase-conjugating mirror and undergo a frequency
shift.
To second order in $L/l$, disorder increases the intensity because it
traps the light near the mirror, where it is
amplified by interaction with the pump beams.
This explains the initial decrease of $\langle R_-\rangle$
followed by an increase in the non-degenerate regime.
The decrease persists in the degenerate regime, because
resonant transmission through the disordered region makes trapping
inefficient.
The resonant transmission is the result of constructive interference of
multiply scattered light, which is made possible by
phase-shift cancellation. 

In conclusion, we have studied the interplay of optical
phase-conjugation and multiple scattering in a random medium. The
theoretical predictions of a reflectance minimum provides a clear
signature for experimentalists in search for effects of phase-shift
cancellation in strong inhomogeneous media. The random-matrix approach
presented here is likely to have a broad range of applicability, as in
the analogous electronic problem\cite{Sto91,sup}. One direction for
future research is to include a second phase-conjugating mirror opposite
the first, with a different phase of the coupling constant. Such a
system is the optical analogue of a Josephson junction\cite{Hout91}, and
it would be interesting to see how far the analogy goes.

This work was supported by the Dutch Science Foundation NWO/FOM.


\begin{references}
\bibitem{voetnoot} Notice that, while ${\cal H}_0$ is Hermitian, the
matrix operator in Eq.~(\protect\ref{BdGlike}) is not. This is an essential
difference with the Bogoliubov-de Gennes operator for Andreev
reflection, which is Hermitian.
\end{references}
\end{document}